\documentclass[aps,showpacs,prl,onecolumn,preprint,letterpaper,nobibnotes]{revtex4-1}
\usepackage{amssymb}
\usepackage{amsmath}
\usepackage{graphicx}
\usepackage{url}
\newcommand\ket[1]{\ensuremath{|#1\rangle}}
\newcommand\bra[1]{\ensuremath{\langle#1|}}
\begin{document}
\title{Carving complex many-atom entangled states by single-photon detection}
\author{Wenlan Chen}
\author{Jiazhong Hu}
\author{Yiheng Duan}
\author{Boris Braverman}
\author{Hao Zhang}
\author{Vladan Vuleti\'{c}}
\affiliation{Department of Physics and Research Laboratory of Electronics, Massachusetts Institute of Technology,
Cambridge, Massachusetts 02139, USA}

\begin{abstract}
We propose a versatile and efficient method to generate a broad class of complex entangled states of many atoms via the detection of a single photon. For an atomic ensemble contained in a strongly coupled optical cavity illuminated by weak single- or multi-frequency light, the atom-light interaction entangles the frequency spectrum of a transmitted photon with the collective spin of the atomic ensemble. Simple time-resolved detection of the transmitted photon then projects the atomic ensemble into a desired pure entangled state. This method can be implemented with existing technology, yields high success probability per trials, and can generate complex entangled states such as multicomponent Schr\"{o}dinger cat states with high fidelity.\end{abstract}
\maketitle

Entanglement is a useful resource in physics. By means of interatomic entanglement, it is possible to overcome the standard quantum limit associated with projection noise in atom interferometers and atomic clocks \cite{squeeze,PRLTak,PRLSmi,PRLIan,JThompson,PNASPolzik,PRLIan2,NaRie,Nematic,NaGross,PRLMitchell}. Entanglement can also be used to implement secure communication networks \cite{DLCZ,internet,NaKu}, and may enable more efficient computation algorithms, potentially with significant impact on computer science \cite{RevModPhysError,Grover,dots,RevModPhysNonAbel,DuanPRL}.

Entanglement in many-particle systems is non-trivial to generate, and often as challenging to experimentally verify. Entanglement implies correlations between the particles, and hence its generation requires controlled interactions between many particles. Therefore, the difficulty of generating entanglement typically dramatically increases both with particle number, and with the complexity of the entangled state. Most entangled states of many atoms generated so far have been relatively simple, characterized by positive Gaussian quasi-probability distribution functions \cite{PRLSmi,PRLIan,JThompson,PNASPolzik,PRLIan2,NaRie,Nematic,NaGross,PRLMitchell,WangXBPR}, or a Wigner function with at most one negative region \cite{NaMc}. More complex entangled states \cite{JMOAgarwal}, Greenberger-Horne-Zeilinger states \cite{GHZ}, have been generated in chains containing up to 14 ions \cite{Leibfried04062004,Roos04062004,PhysRevLett.106.130506}.

It is in general difficult to realize complex non-classical atomic states, as entanglement with unobserved degrees of freedom leads to decoherence. Thus, with the exception of Ref. \cite{NaMc}, all entangled states of many atoms generated to date have been mixed quantum states with low purity. In the work by McConnell et al. \cite{NaMc}, a scheme to generate entanglement in many-atom ensembles by probabilistic photon detection is used \cite{PhysRevA.88.063802,PhysRevA.89.033801,1367-2630-15-1-015002,Luke,Huang}, where probability of entanglement generation is traded in for high state purity on rare but heralded occasions \cite{DLCZ}. However, for that method the success probability decreases exponentially for more complex states with smaller structures in the Wigner function \cite{PhysRevA.88.063802}.

\begin{figure}[htbp]
\begin{center}
\includegraphics[height=1.8in]{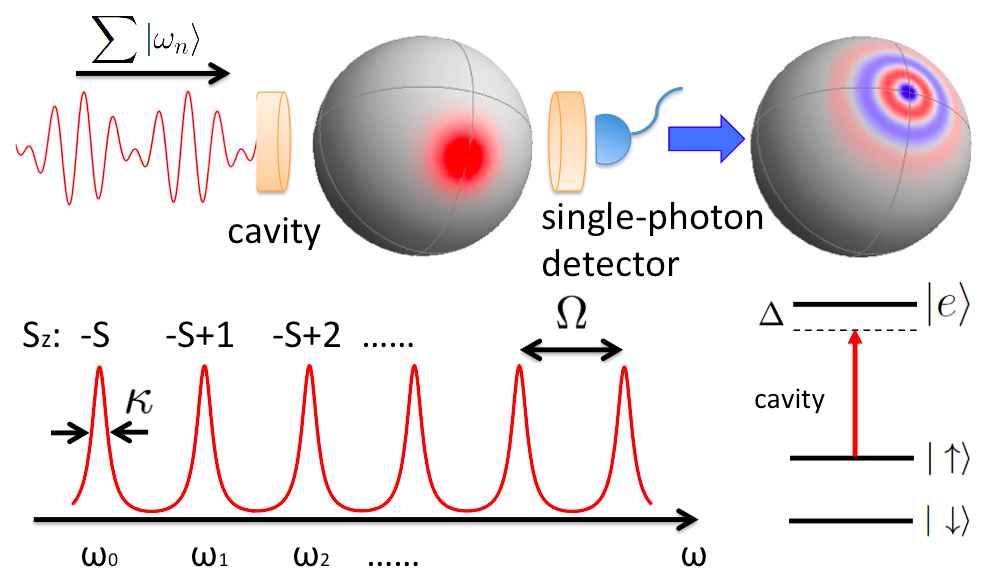}
\caption{Setup for entanglement-carving in atomic ensembles by single-photon detection. The cavity mode couples one of the ground states $\ket{\uparrow}$ to an electronic excited state $\ket{e}$ with a detuning $\Delta$. The system operates in the strong-coupling regime (cooperativity $\eta \gg 1$), such that each atom in $\ket{\uparrow}$ shifts the cavity resonance by an amount $\Omega>\kappa$, where $\kappa$ is the cavity linewidth. We prepare all N atoms in the rotated CSS $\ket{\theta,\phi}$. Then, we send in a weak optical pulse $\sum \ket{\omega_i}$ containing multiple frequency components that coincide with possible cavity resonance frequencies. Once the photon detector registers a transmitted photon, the atomic ensemble is projected into a known entangled state, that is determined by the spectrum of the incoming light and the detection time of the photon.}
\label{fig1}
\end{center}
\end{figure}

In this Letter, we propose a heralded scheme to universally engineer a broad class of complex entangled states simply by the detection of one photon. When a strongly coupled ensemble-cavity system \cite{CQED} is illuminated by a weak light field, the atom-light interaction entangles every eigenstate of the collective atomic spin (Dicke state \cite{PRDicke}) with a corresponding frequency component of a photon transmitted through the cavity. A coherent superposition of different Dicke states with arbitrary amplitudes and phases can be engineered by spectral shaping of the input photon, recording a transmitted photon, and rotating the atomic spin conditioned on the detection time of the transmitted photon. This represents a powerful technique to ``carve" a complex entangled state out of an unentangled product state of the individual atomic spins via the detection of just one photon. The proposed method is efficient in that the generation probability is simply given by the overlap of the initial unentangled state with the target state, divided by the number of frequency components in the incident photon pulse. The accessible states, that include individual Dicke states \cite{bouyer}, squeezed Schr\"{o}dinger cat states, and maximally entangled GHZ-like states \cite{GHZ}, can have small features in phase space, and correspondingly large Fisher information \cite{fisherinfo}, thus enabling atomic clocks and interferometers operating beyond the standard quantum limit \cite{PhysRevA.88.063802}.

We consider $N$ three-level atoms trapped inside, and uniformly coupled to, an optical cavity (Fig.~1). Two ground states, $\ket{\uparrow}$ and $\ket{\downarrow}$, correspond to a pseudospin $\vec{s_i}$ of atom $i$ with $s_i=1/2$, and we define a collective spin $\vec{S}\equiv\sum{\vec{s_i}}$. An excited state $\ket{e}$ is coupled to one of the ground states $\ket{\uparrow}$ by the cavity mode. The detuning between the cavity mode and the atomic transition is $\Delta$. By adiabatically eliminating the excited state, the interaction Hamiltonian can be written as \cite{MonikaPRA}
\begin{equation}
H=\hbar\Omega(S_z+S)\hat c^\dagger \hat c.
\end{equation}
Here, $\Omega=g^2/\Delta$ is the coupling strength, $S=N/2$ is the magnitude of the collective spin $\vec S$, $2g$ is the single-photon Rabi frequency, and for the moment we are ignoring the scattering of photons into free space by the atoms, and the associated reduction in cavity transmission. Each atom in state $\ket{\uparrow}$ shifts the cavity resonance by a frequency $\Omega\ll\Delta$. When $\Omega$ is larger than a few cavity linewidths $\kappa$, each possible value of $S_z$ ($S_z=-S$, $-S+1$, $\ldots$, $S-1$, $S$) corresponds to a resolved cavity line. We define $\omega_c$ as the cavity resonance without any atoms in $\ket{\uparrow}$, so the resonance frequency of the cavity with $n=S_z+S$ (the $n$-th Dicke state) atoms in state $\ket{\uparrow}$ is $\omega_n=\omega_c+n\Omega$.

We first initialize all $N$ atoms in the coherent spin state (CSS) with the polar angle $\theta$ and the azimuthal angle $\phi$:
\begin{equation}
\ket{\theta,\phi}=\left(\cos(\theta/2)\ket{\downarrow}+e^{i\phi}\sin(\theta/2)\ket{\uparrow}\right)^{\otimes N}.
\end{equation}
For the following, it is convenient to write the CSS $\ket{\theta,\phi}$ in the Dicke basis $\ket{S,S_z=-S+n}$. Thus $\ket{\theta,\phi}=\sum_{n=0}^{2S}c_n\ket{S,-S+n}$ with coefficients \cite{Arecchi}
\begin{equation}
c_n=\sqrt{\binom{2S}{n}}e^{in\phi}\cos^{2S-n}(\theta/2)\sin^n(\theta/2).
\end{equation}

We prepare the incident light field by modulating a weak pulse of monochromatic light so that it acquires sidebands at various frequencies $\omega_n$, which coincide with the possible cavity resonance frequencies. The resultant state of the light is expressed as $\ket{\gamma}=\sum_n A_n\ket{\omega_n}$, where $A_n$ is the complex amplitude of frequency component $\omega_n$. When this light is incident onto the cavity, the frequency component $\omega_n$ is transmitted through the cavity only when there are $n$ atoms in state $\ket{\uparrow}$. The transmission of other frequency components corresponds to other values of $S_z$. Thus the strongly coupled atom-cavity system generates correlations between the spectrum of the transmitted light and the possible $S_z$ values of the collective atomic spin. The quantum state of the atom-light system when a single photon has been transmitted is then
\begin{equation}
\ket{\Psi_t}=\sum_{n=0}^{2S}c_n A_n\ket{S,-S+n}\otimes\ket{\omega_n}.
\end{equation}

Subsequently, we measure the transmitted weak light with a single-photon detector. If a transmitted photon is detected in the state $\ket{\gamma'}=\sum_n B_n\ket{\omega_n}$, then the collective atomic spin is projected onto
\begin{equation}
\ket{\psi}=C\sum_{n=0}^{2S}A_n B^*_n c_n\ket{S,-S+n},\label{55}
\end{equation}
where $C$ is a normalization factor.

By controlling the complex coefficients $A_n$ and $B_n$, we can generate an arbitrary quantum state $\ket{\psi}$ of the atomic spin by simply detecting a single photon in the state $\ket{\gamma'}$. The effect is a carving process on the initial CSS. The cavity transmission in combination with photon detection in state $\ket{\gamma'}$ engraves the coefficients of different Dicke states, and projects the collective atomic spin into a chosen, potentially highly entangled state $\ket{\psi}$. The carving is efficient in that the state generation probability is simply given by the projection $\lvert \langle \theta,\phi \vert \psi \rangle \rvert^2$ of the initial CSS $\ket{\theta,\phi}$ onto the desired final state $\ket{\psi}$, divided by the number of frequency components in the incident photon pulse.

\begin{figure}[htbp]
\includegraphics[height=2.8in]{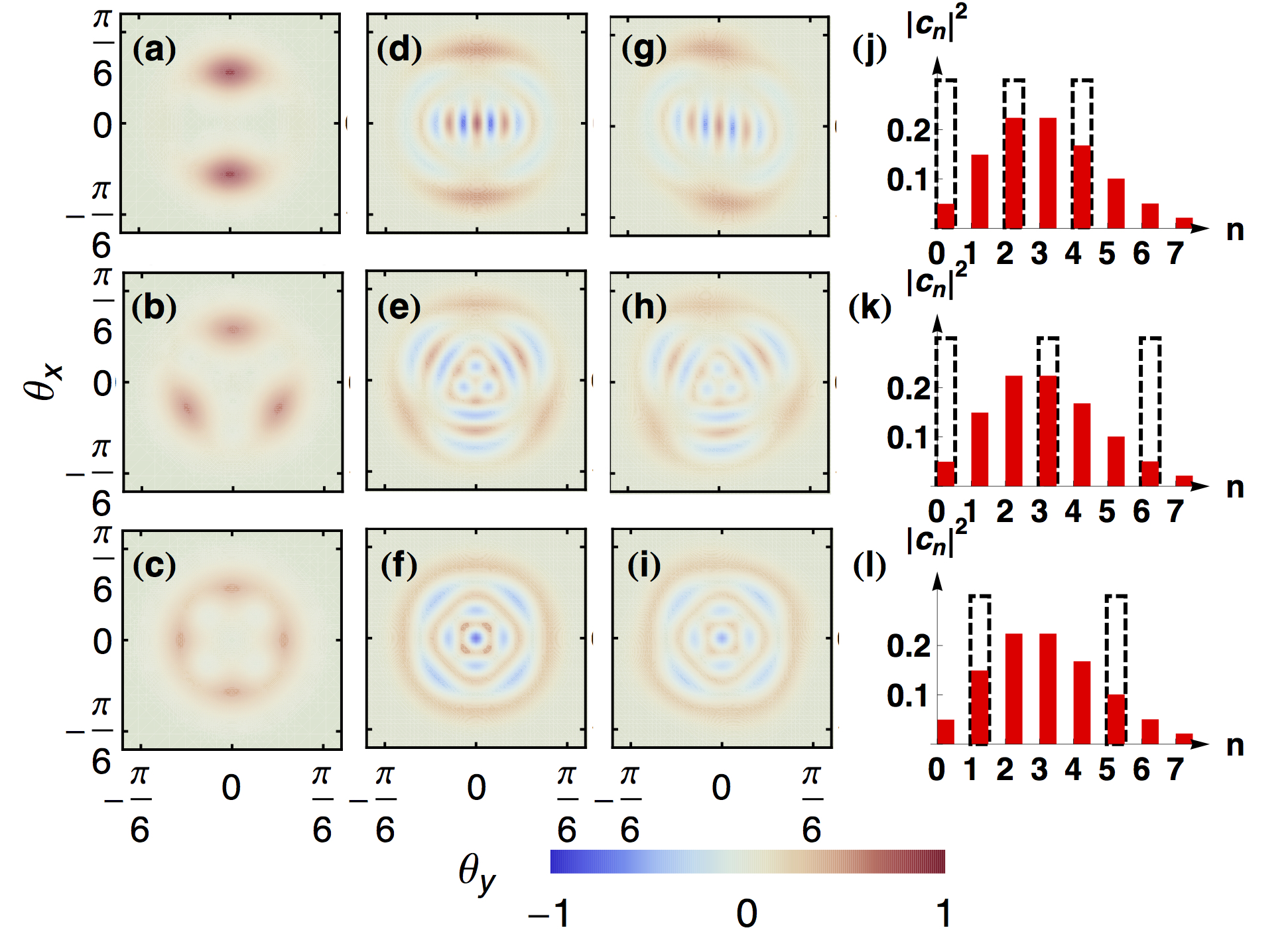}
\caption{Examples of two-, three- and fourfold symmetric cat states of collective atomic spin. 100 atoms are initialized in a rotated CSS $\ket{\theta,\phi}$ with $\theta=0.248$~rad and $\phi=0$. Fig.~(a)-(f) show the entangled states generated with an ideal cavity of infinite cooperativity $\eta$. (a)-(c) are the plots of the Husimi Q function, $Q=2\bra{\theta,\phi}\rho\ket{\theta,\phi}$, where $\rho$ is the atomic density matrix, and (d)-(f) are the plots of the Wigner function. Here each row shows the superposition of Dicke states $n=0,2,4$, $n=0,3,6$, and $n=1,5$, respectively. We also plot the Wigner function for the non-ideal case calculated for $\eta=200$ in Fig.~(g)-(i). Fig.~(j)-(l) show the carving process on the atomic ensemble, with the Dicke state distribution of the initial CSS (red solid bars), and the frequency spectrum of the incident light (dashed black lines).}
\label{fig2}
\end{figure}

We plot a few examples of Schr\"{o}dinger's cat states carved with this method. In Fig.~2, we display both the Husimi Q function and the Wigner function \cite{PhysRevA.49.4101} to characterize the entangled state, where the Q function shows the separation between different quasi-probability regions, and the Wigner function displays the coherence in the form of fringes. Using merely three frequencies, $\omega_q$, $\omega_{q+p}$, $\omega_{q+2p}$, we can generate a $p$-fold symmetric entangled state by projecting the atomic state into different superpositions of $\ket{S,-S+q}$, $\ket{S,-S+q+p}$ and $\ket{S,-S+q+2p}$, where $q$ can be an arbitrary integer.

Since a weak incident light beam can be easily prepared as a superposition of frequency components $\sum_n A_n\ket{\omega_n}$ by a combination of frequency and amplitude modulation, the remaining challenge is how to measure the transmitted photon in the $\ket{\gamma'}$ basis. We propose one universal and simple detection scheme that projects the photon into state $\ket{\gamma'}=\sum_n B_n \ket{\omega_n}$ as below.

Let us start with the simplest case. If the incident single-photon Fock state ($k=1$) or weak light pulse with average photon number $\langle k\rangle\ll 1$ is monochromatic with frequency $\omega_n$, we simply measure the cavity transmission. (We assume that the pulse is long compared to the cavity decay time $\kappa^{-1}$, so that it can be approximated as monochromatic.) If there is a photon detection event, the atomic collective spin is projected into the Dicke state $S_z=-S+n$. This prepares a Dicke state of the atomic ensemble, similar to the scheme of Ref.~\cite{reichel40atoms}, where, however, many photons are used.

If the input pulse corresponds to a superposition of two frequencies, $\ket{\gamma}=A_n\ket{\omega_n}+A_m\ket{\omega_m}$, we record the transmission time $\tau$ of the photon. This situation corresponds to a photon annihilation operator at the detector for time $t$:
\begin{equation}
\hat E^-(t)=A(\hat a_{\omega_n}e^{-i\omega_n t}+\hat a_{\omega_m}e^{-i\omega_m t}),
\end{equation}
where $A$ is a constant coefficient, $\hat a_{\omega_i}$ is the annihilation operator for a photon of frequency $\omega_i$, and we have assumed that the photon detector is sufficiently broadband so that it does not distinguish between the frequency components $\omega_{n}$, $\omega_{m}$. For a given detection time $\tau$, there always exist one ``bright'' photon state $\ket{\gamma_+(\tau)}$ and one ``dark'' photon state $\ket{\gamma_-(\tau)}$
\begin{equation}\label{brightstate}
\ket{\gamma_{\pm}(\tau)}={1\over\sqrt 2}\left(\ket{\omega_n}\pm e^{i(\omega_m-\omega_n)\tau}\ket{\omega_m}\right),
\end{equation}
where $\hat E^-(\tau)\ket{\gamma_-(\tau)}=0$. Once we detect a photon at time $\tau$, it is projected onto the bright state $\ket{\gamma_+(\tau)}$. From Eq.~\ref{brightstate} and Eq.~\ref{55} we see that, for detecting a photon at time $t=\tau$ compared to time $t=0$, i.e., the collective atomic spin is rotated by an angle $-\Omega\tau$ about the $\hat z$ axis. In other words, the detection time $t=\tau$ of the photon simply serves as the new time reference, relative to which always the same entangled state is created.

This scheme also works when the photon has more than two frequency components. For the photon with $p$ frequency components, we write out the corresponding annihilation operator at time $t$ as
\begin{equation}
\hat E^-(t)=A\sum_{j=1}^p \hat a_{\omega_{n_j}}e^{-i\omega_{n_j} t}.
\end{equation}
There is always one bright state,
\begin{equation}
\ket{\gamma_+(\tau)}={1\over \sqrt p}\sum_{j=1}^p e^{i\omega_{n_j}\tau}\ket{\omega_{n_j}},
\end{equation}
and $p-1$ dark states $\{\ket{\gamma_{-j}(\tau)}\}$ for $\tau$, the time that the photon is detected, where $\hat E^-(\tau)\ket{\gamma_{-j}(\tau)}=0$. As before, detection of a photon at time $\tau$, which projects the photon into $\ket{\gamma_+(\tau)}$,  creates the desired entangled state at that time $t=\tau$.

\begin{figure}[htbp]
\includegraphics[width=2.1in]{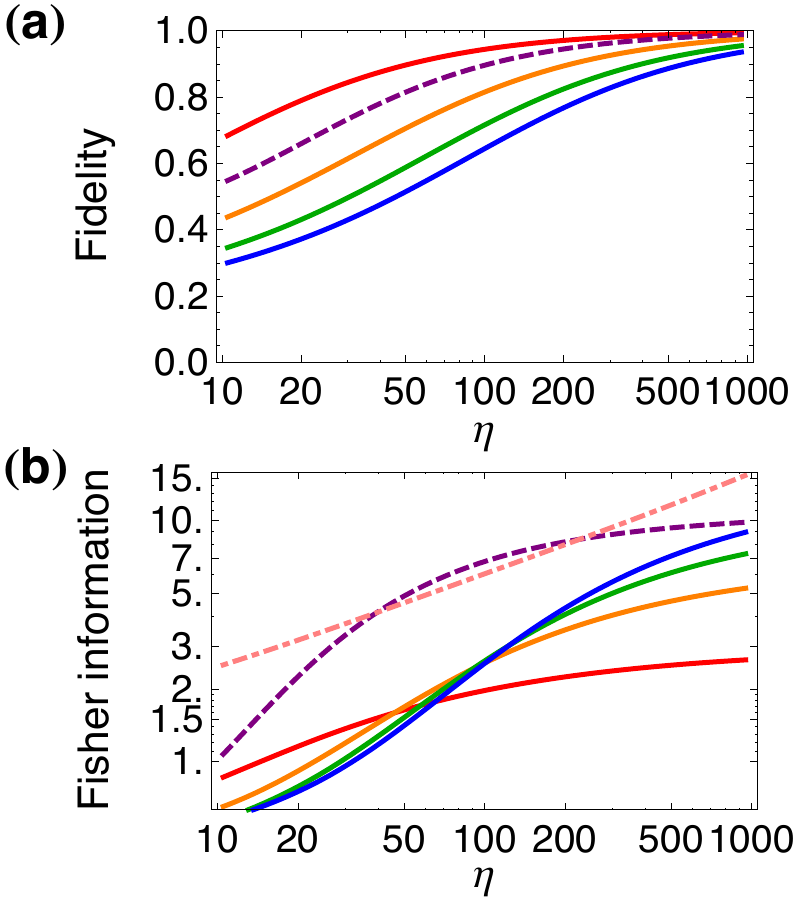}
\caption{(a) Fidelity for generating the Dicke states $S_z=-S+n$ with $n=1,3,5,7$ (solid lines, red, orange, green and blue), and for an equal superposition of the Dicke states $n=1$ and $n=3$ (dashed purple line), for an ensemble of $N=100$ atoms. For a given cooperativity $\eta$, we adjust the cavity-atom detuning and the initial CSS to maximize the fidelity. The fidelity is monotonically decreasing with $n$, due to more spin degrees of freedom involved, and the absorptive broadening of the cavity transmission spectrum. (b) Metrological gain for the same Dicke states and cat state, and for the squeezed state (dot-dashed pink line). This squeezed state is generated by preparing the atomic ensemble in CSS with $\theta=\pi/2$ and $\phi=0$, and sending in a single-frequency pulse $\omega_{N/2}$ which transmits through the cavity at $S_z=0$. The Fisher information, representing the metrological gain, is normalized to the value for the CSS.}
\label{fig3}
\end{figure}

So far, we have assumed that the cavity lines are perfectly resolved, i.e. a photon at frequency $\omega_n$ is transmitted if and only if the atomic collective spin takes on the value $S_z=-S+n$. This situation corresponds to an infinite cooperativity parameter \cite{TanjiSuzuki2011201}. At finite cooperativity, the cavity amplitude transmission function is a Lorentzian of finite width that for $n$ atoms in $\ket{\uparrow}$ given by \cite{TanjiSuzuki2011201}
\begin{equation}
\mathcal{T}(\delta,n)={1\over 1+{n\eta\over 1+4(\Delta+\delta)^2/\Gamma^2}-2 i \left[{\delta\over\kappa}-n\eta{(\Delta+\delta)/\Gamma\over1+4(\Delta+\delta)^2/\Gamma^2}\right]}.\label{10}
\end{equation}
Here, $\eta=4g^2/(\Gamma\kappa)$ is the cavity cooperativity, $\Gamma$ and $\kappa$ are the linewidth of the atomic transition and cavity, $\Delta$ is the cavity-atom detuning, and $\delta=\omega_l-\omega_c$ is the light-cavity detuning. Eq. (\ref{10}) includes the effect of photon emission into the free space that leads to a broadening of the cavity lines. To avoid state deterioration by photons scattered into free space, we limit the input pulse to a single-photon Fock state or a weak coherent state with average photon number $\langle k\rangle\ll 1$. The average transmitted photon number must be kept much smaller than one, in order to avoid that a second undetected transmitted photon creates a different atomic state than the desired state.

We write the transmitted state at finite cooperativity as
\begin{equation}
\ket{\Psi_t}=\sum_{n=0}^{2S}c_n\ket{{S},-S+n} \sum_k A_k \mathcal{T}(\omega_k-\omega_c, n)\ket{\omega_k}.
\end{equation}

When the entangled photon is projected onto state $\sum_k B_k\ket{\omega_k}$ by photon measurement, the atomic state becomes
\begin{equation}
\ket{\psi}=C'\sum_{n=0}^{2S}\sum_k A_k B^*_k \mathcal{T}(\omega_k-\omega_c, n)c_n\ket{{S},-S+n},
\end{equation}
where $C'$ is a normalization factor. The finite cooperativity thus leads to an admixture of Dicke states neighboring the desired Dicke state, and an imperfect spin state fidelity compared to the desired ideal state. We use realistic parameters and plot the Wigner function of the correspondingly generated cat states in Fig.~2(g)-(i), setting $\eta=200$, $\Delta=2\pi\times66$~MHz, $\Gamma=2\pi\times5.2$~MHz, and $\kappa=2\pi\times0.1$~MHz. Compared to Wigner function of the ideal states, the interference structure is unchanged but exhibits reduced contrast. The finite cooperativity also leads to a nonlinear cavity transmission spectrum. It introduces lower transmission and a broader cavity linewidth for higher-order Dicke state component due to larger absorption by more atoms in the $\ket{\uparrow}$ state. The $S_z$-dependent transmission reduction can be compensated by adjusting the input frequency spectrum. However, the broadening of the cavity lines due to atomic absorption deteriorates the cavity filtering behavior, and the admixture of unwanted Dicke states reduces the fidelity of the generated state compared to the target state. The filtering can be improved by increasing the cavity cooperativity $\eta$ and the detuning $\Delta$. This is illustrated in Fig.~3(a), where the fidelity for various Dicke states, and for a superposition of Dicke states, are shown. The state creation probability, when a photon is incident, is between 0.15 and 0.05.

In order to apply this state-carving method for metrology beyond the standard quantum limit, we plot the metrological gain for different Dicke states, for a superposition of Dicke states, as well as for a squeezed state, as a function of cooperativity $\eta$ (Fig.~3(b)). The curves show that the created cat state is robust at finite $\eta$, carrying larger Fisher information than the CSS. For any $\eta$, we can carve a cat state such that it has a larger metrological gain than the spin squeezed state generated by a single photon. The complex states could achieve high metrological gain at a given cooperativity $\eta$.

The cooperativity $\eta$ can be improved by means of micro cavities \cite{microCavity} or higher-reflectivity coatings. If we consider the two-component cat state in Fig~2(a), the predicted fidelity in the experiment should be 0.63 for $\eta=20$, 0.88 for $\eta=100$ and 0.99 for $\eta=1000$. We have verified that this result does not depend on ensemble size.

In conclusion, we propose a universal scheme, that uses only available technology, to efficiently generate a large variety of entangled states by detecting one photon. The method resembles a carving process, which engineers the amplitude and phase of each Dicke state. Variations of this scheme may be used for creating non-classical states of superconducting qubits \cite{SQUID2}, ensembles of quantum dots \cite{dot3}, or mechanical oscillators coupled to a cavity \cite{mech}. 

This work was supported by the NSF, DARPA (QUASAR), and MURI grants through AFOSR and ARO. BB acknowledges support from National Science and Engineering Research Council of Canada. We thank Peter Zoller for discussions and for proposing the term ``quantum state carving".

\bibliographystyle{apsrev4-1}
\bibliography{entangle}
\end{document}